\documentclass[english,aps,prper,reprint,showpacs,titlepage,longbibliography]{revtex4-2}

\usepackage[T1]{fontenc}	
\usepackage[latin9]{inputenc}	
\usepackage{geometry}		
\usepackage{graphicx}
\usepackage[above,below]{placeins}	
\usepackage{times}
\usepackage{multirow}

\usepackage{hyperref}  
\hypersetup{colorlinks=true,urlcolor=blue,citecolor=blue,linkcolor=blue}   
\usepackage{enumitem}        
\setlist{nosep}                 

\usepackage{mathtools}
\usepackage{amssymb}
\usepackage{amsmath}
\usepackage{amsthm}
\usepackage[font=small]{caption}
\usepackage{subcaption}
\usepackage{bbold}
\usepackage{comment}
\usepackage{tabularx}
\usepackage{array}
\usepackage{censor}
\usepackage{cancel}
\usepackage{lipsum, babel}

\newenvironment{iquote}
    {\vspace{-.33\baselineskip}\itshape\list{}{\leftmargin=0.15in\rightmargin=0.15in}%
    \item\relax}
    {\endlist\vspace{-.33\baselineskip}}

\newcommand{\who}[1]{\textbf{#1\emph{:}}}
\newcommand{\etal}{\textit{et al.}}
\newcommand{\Second}{2$^{\text{nd}}$}

\begin{document}

\begin{titlepage}

\title{Investigating graduate student reasoning on a conceptual entropy questionnaire}
\author{Nathan Crossette}
\author{Michael Vignal}
\author{Bethany R. Wilcox}
\affiliation{Department of Physics, University of Colorado - Boulder, Boulder, CO 80309}

\begin{abstract}

Student learning in upper division thermal physics has not been studied to the same extent as in other courses like electromagnetism and quantum mechanics. Studies addressing reasoning and learning at the graduate level are even more limited. In this study, we conducted think-aloud interviews with eight graduate students involving questions centered around a set of entropy related conceptual tasks, two of which are similar to tasks presented to undergraduates in other studies. We discuss patterns in student reasoning on each question then discuss themes that appeared across questions. We identify conceptual resources that students frequently used to reason about the interview tasks and compare them to prior work. We observed graduate students commonly thinking about entropy in relationship to a number of states, even in situations where such a connection was not directly relevant. Graduate students also frequently made direct associations between entropy and temperature, despite there being no general, explicit relationship between the two quantities. On the whole, graduate students demonstrated adaptability and metacognitive awareness in their approach to reasoning about entropy.

\end{abstract}

\maketitle
\end{titlepage}

\section{Introduction \& Background}

Energy and entropy, two core thermal physics concepts, have a wide range of relevancy across the fields of biology, chemistry, physics, and engineering. While energy lacks a formal mathematical definition, it causes little conceptual discomfort among students. Somewhat paradoxically, entropy has a concise mathematical definition:
\begin{equation}
S = k_B \ln W
\label{eq:BoltzmannEntropy}
\end{equation}

\noindent (where $W$ is the multiplicity of a macrostate), yet remains a subtle and difficult to understand concept. Entropy and free energy optimization play a critical role in explaining nearly all molecular biological processes from membrane formation to protein folding to metabolism. The foundational nature of entropy as a physical quantity maximized at equilibrium makes it a powerful and important tool for predicting and understanding the long-term behavior of complicated systems. 

In 2015, Dreyfus~\cite{Dreyfus2015} thoroughly catalogued and summarized prior research on student learning in thermodynamics and statistical mechanics across the disciplines of physics, chemistry, and biology. Current Physics Education Research (PER) literature on upper-division thermal physics is limited, and work specifically related to entropy has mostly explored its thermodynamic and macroscopic contexts such as heat engines and the Carnot Cycle~\cite{Dreyfus2015}. One exception is a study by Leinonen \etal\ which compared the consistency of upper-division undergraduate student reasoning about entropy changes of two objects in thermal contact from both microscopic and macroscopic perspectives. The study found a majority of students applied the \Second\ law of thermodynamics consistently across the problems from both perspectives, though there was a notable minority of students who treated entropy (or, more often, the number of accessible microstates $W$) as a conserved quantity~\cite{Leinonen2015}. This study expanded on research by Christensen \etal, which saw a large number of introductory physics students also treating entropy as a conserved quantity~\cite{Christensen2009}.

Other research on undergraduate understanding of entropy from the thermodynamic perspective has examined student reasoning in various contexts such as ideal gases~\cite{BucyPERC2006} and heat engines~\cite{Smith2015}. A common theme emerging from these studies points to a tendency of students to `over apply' the \Second\ law of thermodynamics to conclude that entropy cannot decrease even locally, which is possibly related to a finding that students sometimes struggle to disentangle systems, surroundings, and the universe~\cite{BucyPERC2006, Christensen2009}. Additionally, a tendency for physics students to neglect to utilize state function properties when reasoning about ideal gasses has been identified~\cite{BucyPERC2006,Smith2015}.

In a study on heat engines and cyclic processes, Smith \etal\ found that students did not articulate the connection between the constraint imposed by the \Second\ law and the Carnot cycle, and a demonstrated some difficulty in distinguishing between differential and net changes of state properties~\cite{Smith2015}. Despite a focus on the macroscopic perspective of entropy, some studies have shed light on student's ideas of the microscopic nature of entropy. In two separate studies, Loverude found that students may struggle to fully differentiated microstates from macrostates, and have not fully connected the idea of entropy with multiplicity~\cite{Loverude2010, Loverude2015}.


The study presented in this article centers around interviews of eight physics graduate students at the University of Colorado Boulder and was intended to probe graduate students' understanding of entropy from both macroscopic and microscopic perspectives. The interview consisted of four physics content questions and one short follow-up discussion question. Two questions, which both addressed the macroscopic perspective of entropy, had been used in prior research to study undergraduates' understanding of entropy. Using these question in interviews with graduate students will allow for a more direct comparison between undergraduate and graduate student reasoning. The two other, new questions address entropy from a more microscopic perspective which has received less attention by prior research. Partial findings from one of our new questions, which presented students with a system of a neutrally buoyant string waving in a water bath, have been reported previously~\cite{Crossette2020}. This problem also directly addressed entropy as it relates to probability, which Dreyfus reports has received little coverage in previous literature~\cite{Dreyfus2015}.

Graduate physics students remain an understudied population within the PER field. Learning more about their conceptual difficulties will indicate truly persistent student struggles, which also provides insight into undergraduate difficulties, and expand the understanding of how graduate students reason and construct models. In addition to potentially improving graduate student learning, future work in this vein will provide more perspective on difficulties experienced by undergraduates and a better understanding of the transition path from novice to expert physicist. Generally, our research goals were to better understand how graduate students reason with entropy, how graduate student reasoning compares with that of undergraduates, and how graduate students apply their understanding of entropy to an unfamiliar situation.

In the following section (Sec.~\ref{Sec:Methods}) we will discuss the methodology of the study, including information on the study participants, development and content of the interview questions, and analysis process. Then, in Sec.~\ref{Sec:Results}, we report student responses and provide analysis to each of the interview questions independently. In Sec.~\ref{sec:Resources}, we analyze themes which appeared across multiple questions in the interview and identify resources students commonly used in reasoning through the interview. We conclude in Sec.~\ref{Sec:Conclusion} with a summary of our results and a short discussion of limitations and future work.


\vspace{-.5\baselineskip}
\section{Methodology}
\label{Sec:Methods}

\subsection{Interview Participants and Format}

Interviews took place in the lead up to and the first several weeks of the Spring 2020 semester (before campus closures due to the COVID-19 pandemic). A total of eight graduate students participated, most of whom $(N = 5)$ were beginning the second semester of their first year. Of the remaining 3 students, one was a second year, one was a transfer student with a Master's degree, and one was in their sixth year. Two of the eight participants were international students. Four students (Beth, Chris, Garth, and Harry) had previously completed a graduate-level course in statistical mechanics and two more (Daana and Erik) were taking a graduate-level statistical mechanics course in the Spring of 2020 at CU Boulder. Seven participants had a physics (or engineering physics) undergraduate background, and one had a background in computer science. All students had taken at least one undergraduate-level thermal physics course except Erik, the student with the computer science background.

An overview of the individual participants is provided in Table~\ref{tab:Interviewees}. Our sample of students roughly reflects the demographics of the graduate student population at the University of Colorado Boulder which is a predominantly white institution that under-represents some demographic groups with respect to the larger populations of both the state of Colorado and the United States. Interviewees were paid volunteers who responded to an email request to the physics graduate student population at the University of Colorado Boulder. All students are referred to by pseudonyms.

\newcolumntype{C}{>{\centering\arraybackslash}X}
\begin{table}[]
\renewcommand{\arraystretch}{1.1}
  \caption{Student participant pseudonyms with brief summaries of their graduate-level standing and whether they self-reported any thermal physics (that is, either thermodynamics or statistical mechanics) related research experience. In the rows for Beth and Erik, MS and CS stand for `Masters of Science' and `Computer Science,' respectively. \label{tab:Interviewees}}\vspace{-\baselineskip}
  \begin{center}
    \begin{tabularx}{\columnwidth}{cCC}
    \hline
    \hline
      Pseudonym & Notes & Thermal Research Experience (Y/N)         \\ 
      \hline
      Alex      & $1^{\text{st}}$ year                   & Y \\ 
      Beth      & Transferred w/ MS                      & N \\
      Chris     & $1^{\text{st}}$ year                   & Y \\
      Daana     & $1^{\text{st}}$ year, international    & N \\
      Erik      & $1^{\text{st}}$ year, CS background    & N \\
      Fred      & $1^{\text{st}}$ year                   & Y \\
      Garth     & \Second\ year, international           & Y \\
      Harry     & $6^{\text{th}}$ year                   & N \\
      \hline
      \hline
    \end{tabularx}
    \end{center}\vspace{-.5\baselineskip}
    
\end{table}

Interviews were conducted in a think-aloud setting~\cite{thinkaloud} where participants were asked to verbalize their thoughts and reasoning as much as possible while working through the questions. The interview questions were printed on LiveScribe paper. In addition to synced written work and audio from a LiveScribe pen, interviews were audio and video recorded. An interviewer (author NC) was present to answer questions, prompt the students to verbalize their thinking, and ask students to further explain their reasoning. The interviewer also asked students a few questions about their research experience, prior thermal physics coursework, and time in the graduate program before participants began the content related questions.

\subsection{Interview Questions}
\label{sec:InterviewQuestions}

\begin{figure}[]\parbox{.9725\columnwidth}{

  \raggedright
  \begin{subfigure}{.232\textwidth}
    \centering
    \includegraphics[width=\textwidth]{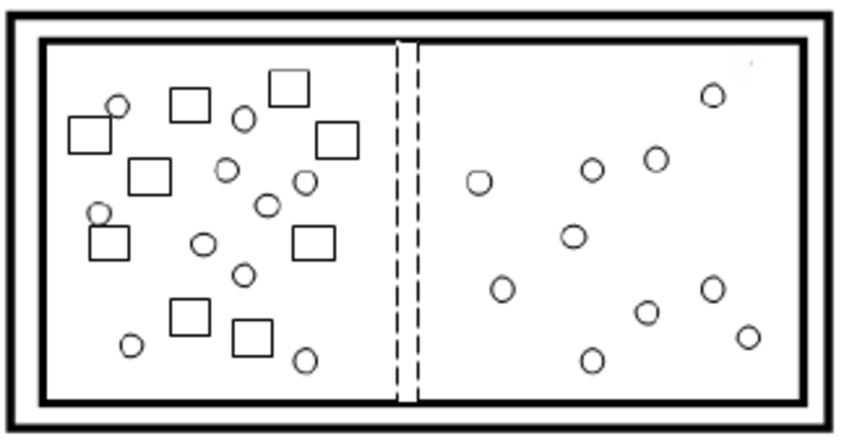}
    \caption{10 circles: 10 circles \label{fig:boxA}}
  \end{subfigure}\hfill
  \begin{subfigure}{.232\textwidth}
    \centering
    \includegraphics[width=\textwidth]{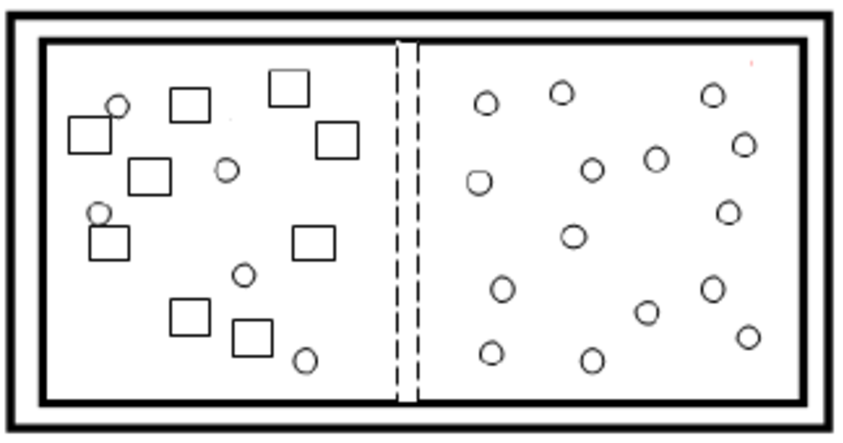}
    \caption{5 circles: 15 circles \label{fig:boxB}}
  \end{subfigure}
  \begin{subfigure}{.232\textwidth}
    \centering
    \includegraphics[width=\textwidth]{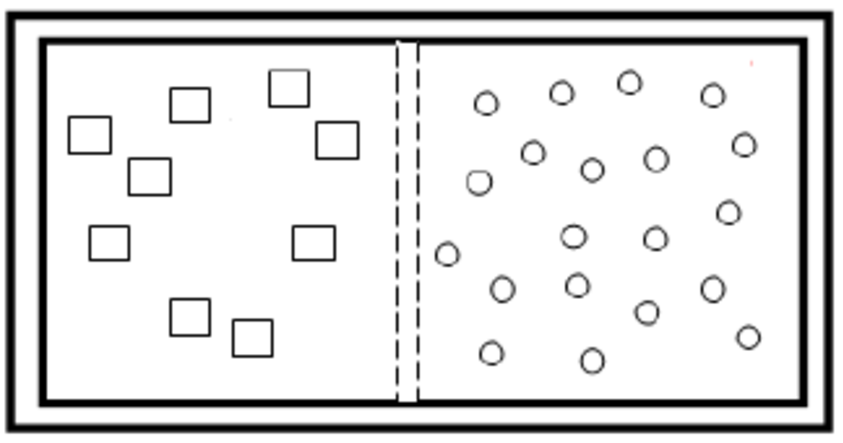}
    \caption{0 circles: 20 circles \label{fig:boxC}}
  \end{subfigure}\hfill
  \begin{subfigure}{.232\textwidth}
    \centering
    \includegraphics[width=\textwidth]{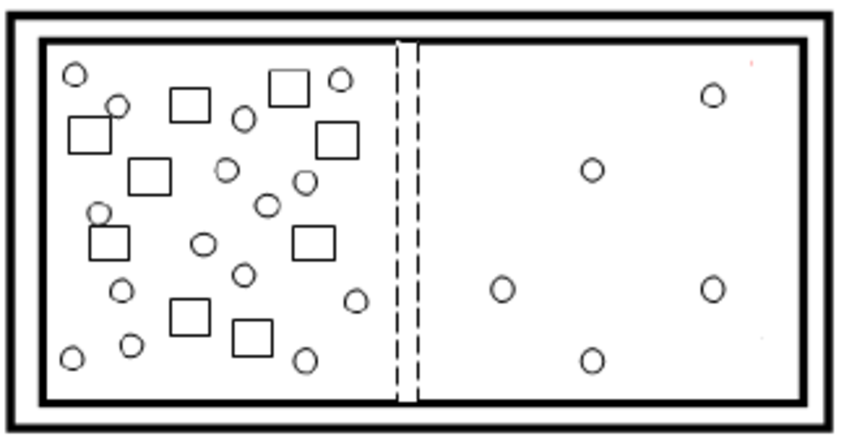}
    \caption{15 circles: 5 circles\label{fig:boxD}}
  \end{subfigure}
  
\begin{itemize}[leftmargin=4mm]
\footnotesize

\item[A)] Rate the states based on their entropy. Which has the highest?

\item[B)] Which of these pictures most closely represents the equilibrium state of the system? How will each state evolve with time?

\item[C)] For each of the four states above, which side (left or right) is at a higher pressure?

\item[D)] Is your answer to the previous question consistent to any claims made in part B about equilibrium? Is anything maximized, minimized or equilibrated at equilibrium? Please elaborate.

\end{itemize}}
  \raggedright
  \caption{\raggedright The Partitioned Box question, the first interview question provided to students. Space was provided for students to show their work.\label{fig:boxes}} \vspace{-1.5\baselineskip}
\end{figure}

The first question, which we will call the ``Partitioned Box'' question, presented students four figures representing different states of a system composed of a box with two sections separated by a semi-permeable membrane and two species of particles: a circle species and a square species. The problem statement told students the circles could freely cross the membrane but the squares could not, and to treat the two species as ideal gasses. These figures and the questions students were asked are shown in Fig.~\ref{fig:boxes}.

Though it was not explicitly stated in the question prompt, temperature could be assumed to be uniform throughout the boxes (if students asked about temperature, they were told to assume it was uniform). Since the temperature and volume of the squares are identical in each of the four states, the entropy of the squares must also be identical in each of the four states. The addition or removal of a circle has no effect on the entropy of the squares since both species are ideal gasses, making the relative entropies of the 4 systems depend completely on the distribution of circles. So while the state depicted in Fig.~\ref{fig:boxB} is a tempting choice for the state with the most entropy --- since the pressures on the two sides of the membrane are the most equlibrated --- the state in Fig.~\ref{fig:boxA} actually has the most entropy. This system is analogous with the osmosis of water across semi-permeable membranes separating two different solutions. Both are examples of a chemical equilibrium where particles are exchanged to equilibrate the chemical potential of each species across the membrane. For mechanical equilibrium to be established, where pressure equalizes, the systems must be able to exchange volume.

By asking first about entropy and equilibrium, we wanted to see the first, `instinctual' method with which students rank entropy. Then, by asking about pressure, we wanted to see how they would contend with thinking about a property (pressure) that could mislead a student by suggesting a consistent, though incorrect, set of answers to this first question: that the state with the most equal pressures between the left and right side of the box has maximum entropy and therefore the equilibrium state, and that the pressure difference is minimized at equilibrium.

The second interview question, which we will call the ``Blocks'' question, was meant to be more familiar to students, and is similar to one used by Christensen \etal \cite{Christensen2009} and Leinonen \etal \cite{Leinonen2015}, but was developed independently. It presented students with an isolated system of two solids (Block A and Block B). Initially block A was at a higher temperature than block B, then the two blocks are placed into thermal contact. We then asked what happened to the blocks, and what could be said about the entropies of the individual blocks before and a long time in the future.

This question was intended to provide interviewees with a bit of a reprieve in between the more difficult first and third questions, but also to probe to what extent students consider entropy to be a `substance' that can be transferred from one object to another. Because nothing was specified about the volume, mass, and heat capacities about the two blocks, nothing can be said about the entropies of the blocks before and after except that Block A's entropy decreased, and Block B's entropy increased by an amount more than the decrease in Block A.

\begin{figure}[]\parbox{.9725\columnwidth}{

  \centering
  \begin{subfigure}{.15\textwidth}
    \centering
    \includegraphics[width=\textwidth]{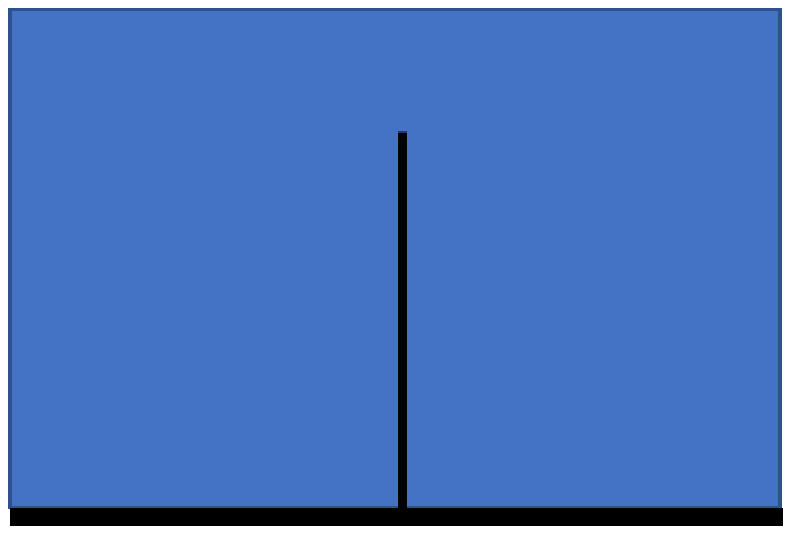}
    \caption{\label{fig:stringA}}
  \end{subfigure}
  \begin{subfigure}{.15\textwidth}
    \centering
    \includegraphics[width=\textwidth]{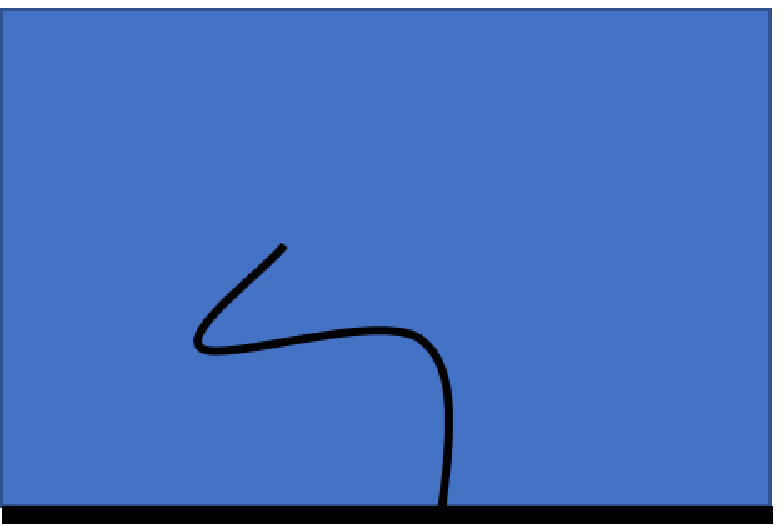}
    \caption{\label{fig:stringB}}
  \end{subfigure}
  \begin{subfigure}{.15\textwidth}
    \centering
    \includegraphics[width=\textwidth]{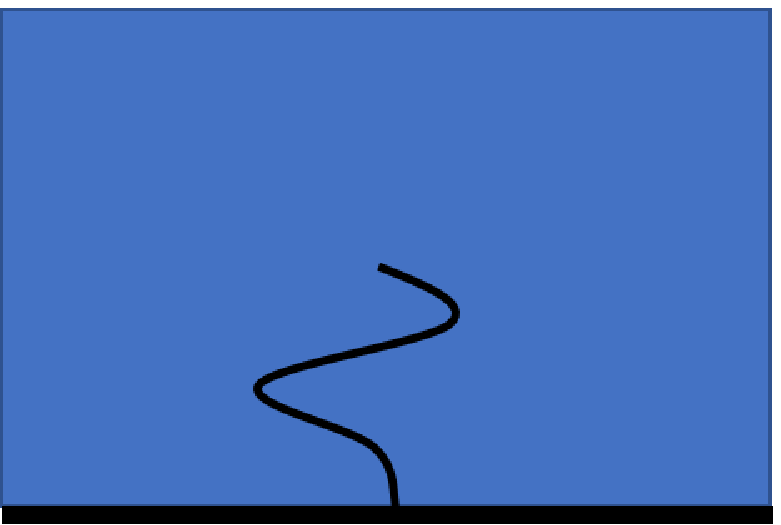}
    \caption{\label{fig:stringC}}
  \end{subfigure}

\begin{itemize}[leftmargin=4mm]
\footnotesize

\item[A)] Based on your intuition, rank the probabilities of finding the string in each of the three conformations shown above.

\item[B)] Is there a property of the string that can be used to define a set of distinct macrostates of the string? Are the conformations shown in the figure macrostates or microstates?

\item[C)] Based on your answer to part B, how would you rank the probabilities of finding the strand in each of the three conformations above?

\item[D)] Can you discuss what is meant by the ``entropy'' of the string, and how it relates to the possible conformations of the string?

\end{itemize}}

  \caption{The first part of the Strings question provided to students. Space was provided for students to show their work.\label{fig:strings}} \vspace{-1.5\baselineskip}
\end{figure}

The third question, the ``Strings'' question, centered around a simplified biophysical system, likely unfamiliar to most physics students: a neutrally buoyant string waving in a water bath. The string was attached at one end to a wall of the water bath. The first part of the question started with three snapshots of different string conformations (or arrangements) and asked students about probabilities, microstates, macrostates, and entropy (see Fig.~\ref{fig:strings}). The key to this question was to recognize each of the three snapshots as microstates which means the probabilities of finding the system in each of the states are all equal. In terms of defining macrostates, we did not consider there to be only one correct answer. As long as the classification property provided by the student generally outlined a means of partitioning microstates, we considered their response to be correct. On this task, we were more interested in the range of definitions generated by students. We reported results of the analysis of student responses to this first part of the Strings Question in a prior article~\cite{Crossette2020}.

\begin{figure}[]\parbox{.9725\columnwidth}{

  \centering
  \begin{subfigure}{.4\textwidth}
    \centering
    \includegraphics[width=\textwidth]{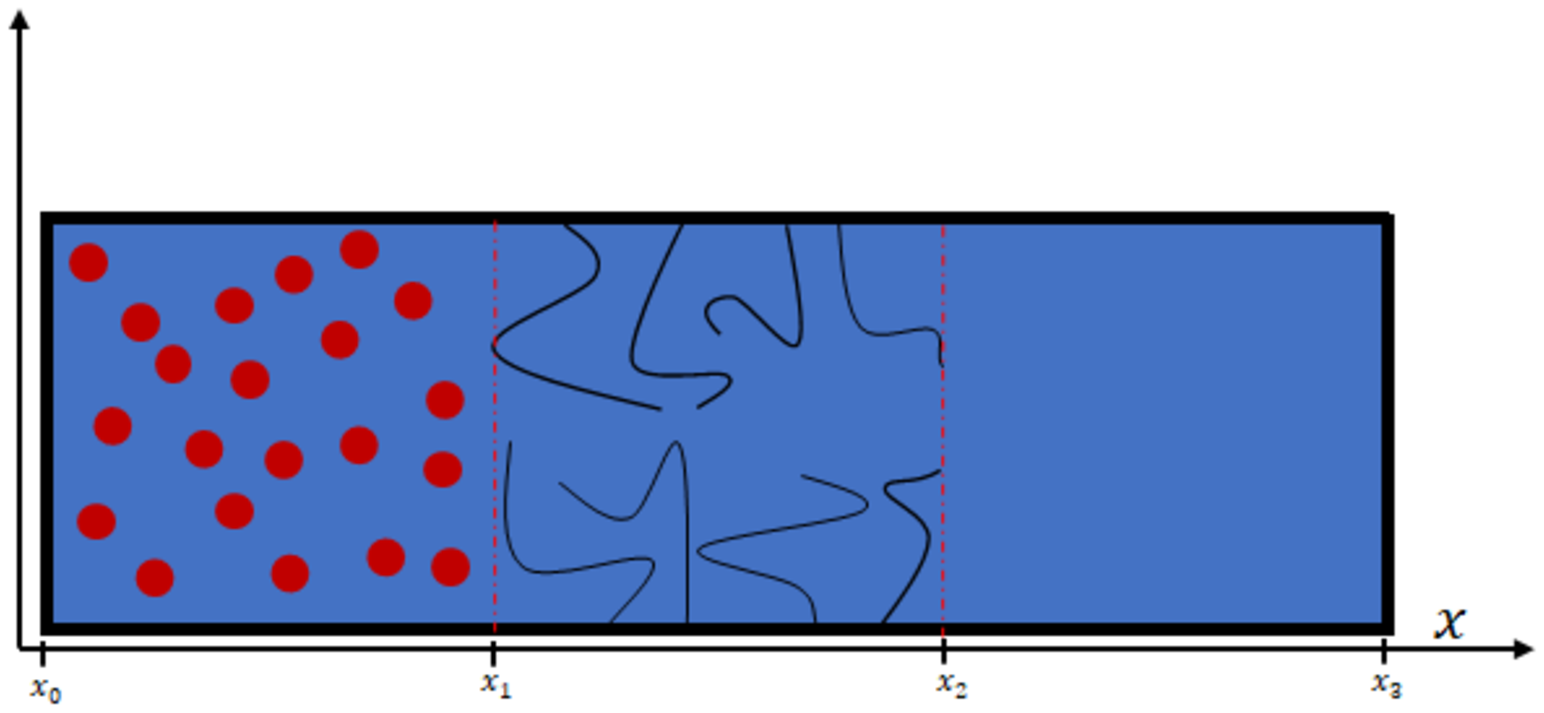}
    \caption{\label{fig:channel}}
  \end{subfigure}
  \begin{subfigure}{.45\textwidth}
    \centering
    \includegraphics[width=\textwidth]{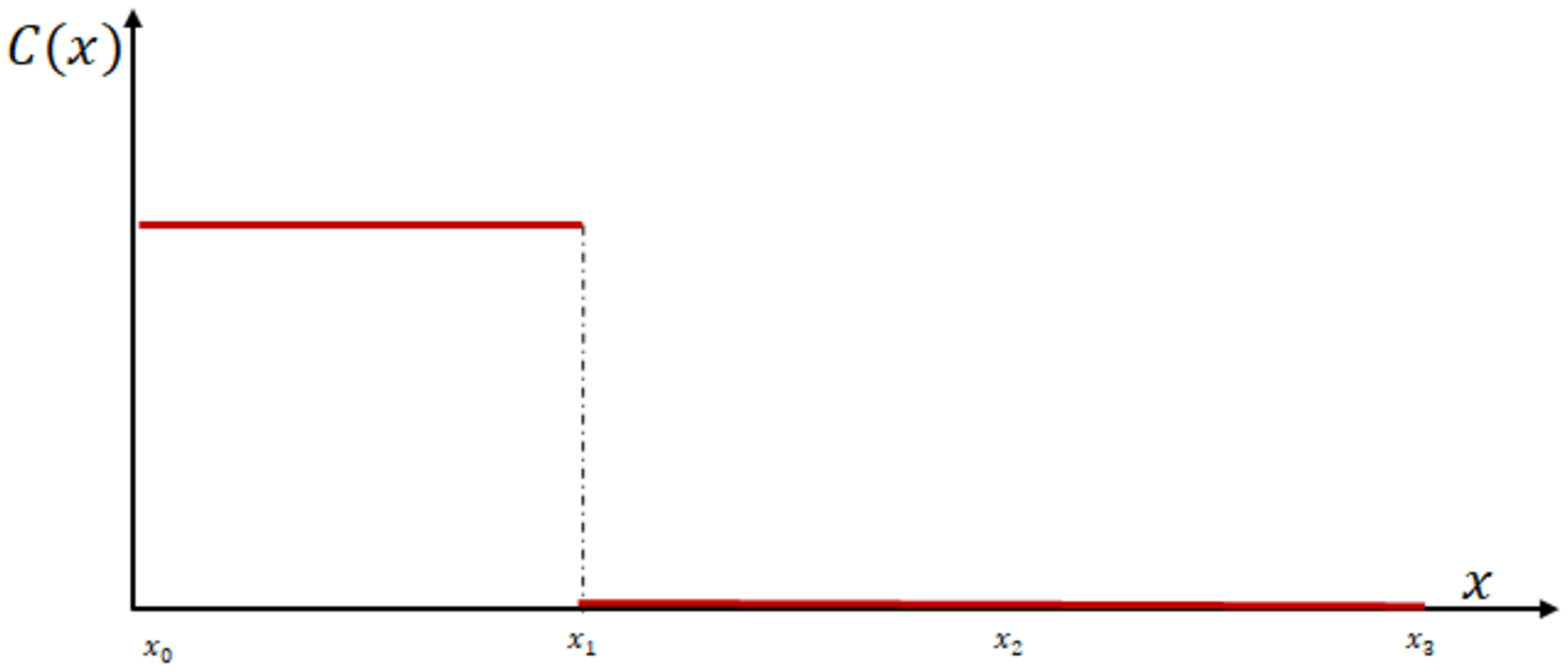}
    \caption{\label{fig:profile}}
  \end{subfigure}

\begin{itemize}[leftmargin=4mm]
\footnotesize

\item[i)] What happens to the number of possible conformations of the strings when the molecules enter the region with the strings?

\item[ii)] What will the concentration profile, as a function of the distance along the channel, of the molecules be after a long period of time? Explain your reasoning.

\end{itemize}}

  \caption{The second part of the String question provided to students. Space was provided for students to show their work.\label{fig:ChannelProfile}} \vspace{-1.5\baselineskip}
\end{figure}

The final part of the Strings question, Part E, took a deeper step into the string system and was not discussed in our prior article. We presented the students with a new, more complicated system of a channel with three sections and with strings attached to the walls in the middle section (Fig.~\ref{fig:channel}). The first and third sections were free of strings, but the first section was filled with red circular particles (balls) suspended in the water. We asked students what happened to the entropy of the strings when a ball entered the region with the strings, then to draw a plot of the concentration profile of the balls throughout the channel after a long period of time had passed.

This question was structured to guide students' reasoning surrounding the plot of the concentration profile by first having them think about the `entropic cost' of having balls in the region with the strings. A ball entering the strings region between $x_1$ and $x_2$ in Fig.~\ref{fig:ChannelProfile} decreases the available conformations of the strings, and therefore reduces the entropy of the strings. However, having the balls diffuse through the system increases the entropy of the balls, so there will be a trade-off between the increase in entropy of the balls and a decrease in the entropy of the strings, resulting in a symmetrical concentration profile with a dip in the center region.

\begin{figure}
  \centering
  \includegraphics[width=.9725\columnwidth]{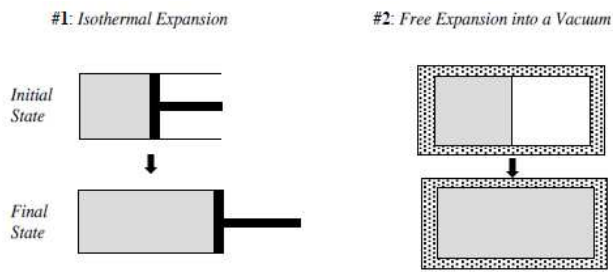}
  \caption{\raggedright Diagram provided to students in the prompt of the Expansions Question. This figure was taken from Bucy~\cite{BucyPERC2006}.\label{fig:expansions}} \vspace{-1.5\baselineskip}
\end{figure}

The final physics-content question, which we call the ``Expansions'' question, was taken nearly in its entirety from a study done by Bucy \etal \cite{BucyPERC2006}. In this question, students were shown diagrams of two ideal processes: an isothermal expansion and a free expansion into vacuum. Both gases began at the same volume, pressure, and temperature and expand to the same final volume (see Fig.~\ref{fig:expansions}). The first gas, undergoing the isothermal expansion, is in contact with a reservoir (necessary to keep the temperature constant) and the second gas is completely isolated from the outer environment. Students were asked about the signs of the changes in entropy of the two gases, to compare the magnitudes of the changes, and to compare the changes in entropy of the outer environments of the two gases. This final task was a small extension of the question from Bucy's study which was added due to the findings of Christensen~\cite{Christensen2009, BucyPERC2006} which found that undergraduates struggle to disentangle the difference between a system, its surroundings, and the universe.

As it turns out, the two gases both increase in entropy by the same amount. The first gas's temperature is constant, so its internal energy, $U = \frac{3}{2}N k_B T$, remains constant. The second gas does no work, and exchanges no heat with its environment, so its internal energy remains constant as well. Since the two gases also end at the same volume, it must be true that the two gases are in the same final state because their volumes and temperatures, two of their state variables, are identical. The two processes differ in that the isothermal expansion is a reversible process, so the increase in entropy of the gas is canceled by the decrease in entropy of the environment. The second process is irreversible so the total change in entropy of the gas and the outer environment is positive.

After completing the questions described above, students were asked to discuss their conception of entropy as a short follow-up to the interview.

Before interviews were conducted, these questions were vetted by thermodynamics instructors from Chemistry, Engineering, Biology, and Chemical Engineering. The questions were also piloted in two interviews: one with a content expert, and one with a PER graduate student.

\subsection{Analysis}

To analyze student responses, written drafts of the interviews were obtained from Otter.ai transcription software. The drafts were then manually checked and edited using the recorded video. Interview transcripts were emergently coded based on student responses to question prompts. Our coding methodology was inspired by Hammer's resources framework~\cite{Hammer}, and thus focused on categorizing common responses and reasoning elements to identify patterns and themes in students approaches and answers to each question. After initial coding by the first author, all three authors collaboratively reviewed sections of interviews to verify code assignments and reach agreement on difficult-to-code passages.

The interviews were coded and studied question by question to break the analysis up into more easily processed pieces. Summaries of student responses to each question were produced which allowed for the easy identification the range, commonalities, and differences of responses and reasoning both within and across questions. Approaching the analysis with the resources framework allowed for inter-question patterns in student reasoning to emerge from the coded transcripts and summaries. Additionally, this approach allowed for a more direct comparison with a study by Loverude~\cite{Loverude2015} which identified entropy-related resources used by undergraduate students. We have a deeper discussion of the definition of resources used in this study at the beginning of Sec.~\ref{sec:Resources}.

\vspace{-.5\baselineskip}
\section{Results \& Discussion of Individual Questions}\label{Sec:Results}
This section will summarize the responses given to each of the four physics-content questions. We will provide some analysis of individual questions here, but a deeper analysis of themes seen across the entire interview will be left for the following section (Sec.~\ref{sec:Resources}).

\subsection{Question 1: Partitioned Box}

Students' ranking of the four states of the system by entropy are summarized in Table~\ref{tab:Q1_EntropyRankings}. Five out of the eight students settled on the correct ranking (for ideal gasses) of $A>D=B>C$ by the end of the question, and seven out of eight correctly identified state $A$ as having the greatest entropy. Since the question involved ideal, non-interacting gasses rather than real gasses, the entropy of states $D$ and $B$ are identical. However, in the case of real gasses, whose particles take up volume, case $D$ would have a negligibly lower entropy than case $B$ due to a volume exclusion effect of real gas particles. In case $D$, the excess of particles on the left results in fewer positional microstates available to the system, since two (non-ideal) particles cannot occupy the same space. Therefore the entropy of the system of \textit{real} gasses increase (negligibly) by moving to a macrostate with fewer circles on the left. This sense, coupled with an intuition to have the pressure difference between the two sides minimized at equilibrium, led us to expect the dominant misconception seen in this question to be ranking the entropy of state $B$ to be higher than the entropy of state $D$.

\renewcommand{\c}[1]{\multirow{2}{*}{\vspace{1pt}#1}}
\begin{table}[t]
\renewcommand{\arraystretch}{1.1}
  \caption{Summary of student entropy rankings for part A of the first question \ref{fig:boxes}. All the rankings generated by students are listed in the leftmost column.  \label{tab:Q1_EntropyRankings}}\vspace{-\baselineskip}
  \begin{center}
    \begin{tabularx}{\columnwidth}{cCC}
    \hline
    \hline
      Ranking   & Initial Ranking & Final Ranking                  \\ 
      \hline
      \c{$A>D=B>C$}      &  Chris, Erik, Garth,   &  Chris, Erik, Garth,           \\
                         &  and Harry                &  Harry, and Daana                \\
      $A>D>B>C$      & Fred and Daana\footnote{given as intermediate ranking between initial and final answers}        & Fred             \\
      $D>A>B>C$      & Beth and Daana &  -                            \\
      $A>B>D>C$      & -    & Beth                              \\
      $B>A>D>C$     & Alex &  Alex               \\
      \hline
      \hline
    \end{tabularx}
    \end{center}\vspace{-1\baselineskip}
    
\end{table}

The graduate students had relatively little trouble identifying states $A$ and $C$ as the states with the most and least entropy, respectively, but had more trouble determining where to rank states $D$ and $B$. Most students ($N=5$) gave the correct ranking, but interestingly, the second most common ranking given by graduate students was to rank state $D$ as having a larger entropy than state $B$, with three students giving this form of ranking a total of four times at some point in the interview. An appeal to mixing was the reason leading Fred and Daana for the entropy of state $D$'s being greater than the entropy of state $B$:

\begin{iquote}
  \who{Daana} The ones which have a more even mixture of squares and circles should have a higher entropy. And, uhh, ones which are more uniform, that is like, if they have all circles or all squares, those should have less entropy.
\end{iquote}

\noindent Fred had a similar sense that seemed to indicate an intuition that entropy is greater the closer the particle ratio is to 50:50 on the left side of the partition.

Slightly less prevalent were rankings with state $B$ having greater entropy than state $D$, with two students (Alex and Beth) giving this ranking at some point in their interviews. Beth gave this ranking initially, before re-ranking at the end of the question after recognizing state $A$ as equilibrium and a prompt from the interviewer which asked whether anything was maximized, minimized, or equilibrated at equilibrium. Interestingly, Alex appealed to mixing as justification for the ranking of $B>A>D>C$, demonstrating that terms used to describe entropy like ``mixing'' mean different things to different people.

Graduate students utilized a variety of terms as descriptive proxies for entropy in the context of this question. The most common terms had to do with `mixing' or that entropy had to do with how the particles were distributed. Previous research has found that even when the term `disorder' is avoided in the physics classroom, undergraduate students still frequently use the term as a resource for thinking of entropy~\cite{Loverude2015}, possibly due to encountering the term in prior high school or college level science courses. Interestingly, explicit use of terms like `order/disorder' only came up with two students on this question, possibly indicating that the term loses favor as a moniker for entropy as physics students advance to the graduate level. Other proxy terms and definitions included connections between entropy and volume/spreading out, the number of microstates, information, and `the thing maximized in the long run.'

Common among the students giving the correct ranking was a strategy of ignoring the square particles for the purposes of ranking entropy. This indicates a competency among the graduate students to simplify problems to the most salient form and understand that the entropies of ideal gasses are additive. Fred was the only student who articulated this reasoning and did not give the $A>D=B>C$ ranking. For Fred, the presence of the squares in a sense, broke a symmetry through some `mixing' mechanism:

\begin{iquote}
  \who{Fred} If the squares were irrelevant, then [states $B$ and $D$] would be equal... I understand that mixing two gases across a membrane should increase entropy... I think I also have to consider the volume not just mixing, so that's why I went with $A$ for the first. But I guess, intuitively, it's more mixed in $D$ than $B$, umm, so I guess if I'm considering mixing and the squares to be relevant than I would say $D$ before $B$ because there are more circles mixed in in $D$ than $B$.
\end{iquote}

The graduate students in the sample had no trouble ranking the pressures on either side of the partition. Most appealed to the ideal gas law, though one student uniquely reasoned through it by thinking about what the particle flux of the circles would be. When it came to confronting whether their pressure rankings were consistent with their choice for equilibrium, three students recognized and explicitly stated that the pressure would not necessarily equilibrate between the two sides of the box. Two students experienced a little dissonance, but did not change their choice of equilibrium and accepted that the left and right pressures would be unequal at equilibrium. One of the students concisely articulated this discomfort:

\begin{iquote}
  \who{Harry} I guess it does seem kind of strange to me if I'm claiming that [state] $A$ represents the, umm, equilibrium case, it does seem a bit odd to me as I'm looking at it now that there would be an unequal pressure, it feels like those should be... should cancel out. But I guess there is also a gradient in chemical potential between the barriers... Umm, so, I'm not exactly sure how to think about that.
\end{iquote}

Three students (Alex, Beth, and Fred) explicitly stated that the left and right pressures should be equilibrated at equilibrium, which is interesting since both Beth and Fred choose state $A$ as the equilibrium state. Beth later changed her ranking (from $D>A>B>C$ to $A>B>D>C$), but not until prompted for an answer about what would be maximized or minimized at equilibrium. Fred admitted that his pressure rankings were not consistent with his choice for equilibrium and that state $B$ would look more like equilibrium, but then expressed an uncertainty with the relationship between entropy and pressure. For Alex, the pressure ranking turned out to be consistent with his choice of state $B$ as the equilibrium state.

Additionally, while reasoning about the pressures, two students considered whether pressure depended on the masses of the particle species, which has been noticed in prior work on undergraduate understanding of the ideal gas law~\cite{Kautz-Micro2005}.

\subsection{Question 2: Blocks}
\label{sec:Q2_Blocks}

Much of the previous research examining student reasoning on two-object systems have found that some undergraduates inappropriately try to conserve entropy, or over apply the \Second\ law by thinking entropy must increase everywhere~\cite{Leinonen2015, Christensen2009, Loverude2015}. Our framing of this task specifically cued students to think about the two blocks separately and did not make any statements about the relative sizes of the two blocks. Our question is quite similar to the macroscopic question from the diagnostic test from the study by Leinonen \etal, but was intentionally less structured and posed to graduate students in a think-aloud format.

In their responses to this question, all interviewees correctly stated something about how the temperatures of the two blocks will equalize to a common temperature. If this was not the first thing they considered, a statement about the temperatures occurred very early in their reasoning about the question.

When considering the entropies of the two blocks, six out of the eight students began by claiming that total entropy of the combined two-block system would increase, possibly as a way of getting oriented and grounding themselves. Notably, none of them explicitly mentioned the \Second\ law by name.

\begin{iquote}
  \who{Beth} The entropy in the final state will be higher than it was in the initial state.
  \who{Interviewer} The entropy of what will be higher?
  \who{Beth} The entropy of this system.
\end{iquote}

\begin{iquote}
  \who{Harry} And the individual entropies... So the entropy of the entire system... the entropy is got to increase.
\end{iquote}

\noindent A seventh student, Alex, began thinking about the total entropy before trailing off then providing answers for the individual blocks with out making any statement about the total entropy. Interestingly, Alex was the only student who explicitly claimed that the changes in entropies of the two blocks would be equal and opposite, implying a `conservation of entropy' similar to that identified by Leinonen \etal. The eighth student, Garth, reasoned directly about the individual entropies by relating entropy to temperature. We will discuss this association in more depth later in this section (see quote from Garth below).

This pattern of first considering the total entropy was one of the more striking observations of this task, particularly since the question prompt asked students to reason about the entropy changes of the individual blocks and that students naturally and effectively thought about constituents of a system (the two gasses) independently in the previous problem. Furthermore, the \Second\ law does not provide insight into the changes in entropy of the constituents of a system, so students had to turn towards other means of reasoning about the entropies of the two blocks.

A common alternative mode of reasoning about the entropies of the blocks was through an association between entropy and temperature. Garth mentioned this idea early on in the question when considering the initial entropies of the two blocks:

\begin{iquote}
  \who{Garth} If something is at a higher temperature, its atoms are vibrating more. So it should have greater entropy...
\end{iquote}
 
\noindent and also as one of his final thoughts after realizing there wasn't enough information to know the initial entropies of the blocks:
 
\begin{iquote}
  \who{Garth} So, I'm definitely associating entropy with temperature and I think if [block] A is cooling than its entropy should be less, and if [block] B is heating up its entropy should increase.
\end{iquote}

Four students made a connection that higher/increasing temperature means more/increasing entropy, as if there were a monotonic function relating entropy and temperature. Such intuition that entropy is functionally related to temperature is not poorly grounded since this connection is consistent with the fact that the entropies of Blocks A and B will decrease and increase, respectively. Though it lead some students to the correct answer, this intuition oversimplifies the true nature of the process: the sign of the change in entropy, $\Delta S = Q/T$, depends on whether the heat flow, $Q$, is positive or negative.

A positive heat flow often results in increases in both temperature and entropy, but overlooking heat flow and considering temperature instead can lead to incorrect conclusions. For example, in spontaneous, endothermic, chemical reactions the heat flow into the system is positive, but the temperature can decrease while entropy increases. In the Appendix, we will discuss, in-depth, a process in the familiar context of ideal gasses that also disproves a simple connection between entropy and temperature. The process involves a non-quasistatic expansion of an insulated gas in which the gas does work causing it's temperature to decrease, while an increase in volume increases entropy more than the drop in temperature reduces entropy.

A non-quasistatic process, something which generally is not covered deeply in thermal physics courses, was required for this counterexample. The little emphasis placed on non-quasistatic processes, along with the fact that for most cases the heuristic that increasing temperature increases entropy is true leaves little mystery why students would lean on this association. However, two students (Alex and Fred) who made this connection between entropy and temperature used to reason that the initial total entropy of Block A was greater than the total entropy of Block B.

\begin{table*}[htbp]
\renewcommand{\arraystretch}{1.1}
  \caption{Summary of graduate student responses to parts A, B, and C of the strings question (third question of the interview). Letters in parentheses on title row refer to prompts in Fig.~\ref{fig:strings}. Bold entries in the final column indicate rankings that changed after students considered macrostates and microstates. \label{tab:stringspt1}}\vspace{-\baselineskip}
  \begin{center}
    \begin{tabularx}{\textwidth}{ccCCc}
    \hline
    \hline
      Student & Initial Ranking (A)& Consideration in Initial Ranking (A) & Ideas for Macrostates (B) &  Final Ranking (C)\\ 
      \hline
      Alex      & $P(c) > P(b) > P(a)$      & deviation from center                     & net deviation, conformations          & $\boldsymbol{P(a) > P(c) > P(b)}$     \\\vspace{-3pt}
      \c{Daana} & \c{$P(c) = P(b) = P(a)$}  & fundamental assumption of                 & \# of turns, net deviation,           & \c{$P(c) = P(b) = P(a)$}  \\
                &                           &statistical mechanics                      &`range' between points                 &                           \\\vspace{2pt}
      Beth      & $P(c) > P(b) > P(a)$      & `wavyness'                                & \# of turns                           & $\boldsymbol{P(c) = P(b) = P(a)}$      \\\vspace{-3pt}
      \c{Erik}  & \c{$P(c) = P(b) = P(a)$}  & there is no preference for                & \c{conformations}                     & \c{$P(c) = P(b) = P(a)$}  \\\vspace{2pt}
                &                           &a particular conformation                  &                                       &                           \\
      Chris     & $P(c) > P(b) > P(a)$      & `wiggliness', amount of deviation         & distance between ends                 & $\boldsymbol{P(c) = P(b) = P(a)}$      \\\vspace{-3pt}
      \c{Fred}  & \c{$P(a) > P(c) > P(b)$}  &\c{net `closeness' of string to center}    & net deviation, conformation,          & \c{$P(a) > P(c) > P(b)$}  \\\vspace{2pt}
                &                           &                                           & \# of turns                           &                           \\
      Garth     & $P(c) > P(b) > P(a)$      & distance between string ends              & distance between ends, position of end& $\boldsymbol{P(c) = P(b) = P(a)}$      \\\vspace{-3pt}
      \c{Harry} & \c{$P(c) = P(b) = P(a)$}  & fundamental assumption of                 & distance between ends,                &\c{$P(c) = P(b) = P(a)$}   \\
                &                           &statistical mechanics                      & net deviation, \# of turns            &                           \\
      \hline
      \hline
    \end{tabularx}
    \end{center}\vspace{-1.5\baselineskip}
    
\end{table*}

In total, five students eventually decided that the entropy of Block A (the hotter block) would decrease and that the entropy of Block B (the colder block) would increase. A sixth student, Erik, initially stated, with some reservation, that the entropy of both blocks would increase. He then nearly corrected himself after considering the equation $T = dU/dS$, but did not express more confidence in this reasoning than his previous reasoning since he could not identify the flaws or veracity of either argument. 

The idea that entropy of both blocks increased came up in Fred's interview as well. Fred also employed the `higher temperature, more entropy' argument to say that the entropy of Block A was initially higher, but that the two blocks would ``approach'' the same entropy once they reached thermal equilibrium, which seems to contradict the claim that both blocks' entropies increased and the `higher temperature, more entropy' heuristic. Fred admitted to thinking of the two blocks as identical, but did not change his answer after considering that the two blocks might be different.

Two students considered whether this process might be isentropic/reversible after stating that, generally, total entropy must either increase or remain constant:

\begin{iquote}
  \who{Erik} A long time after they came in thermal contact, the total entropy of the system will be higher than it was before, because that's how entropy works. Unless... well, greater than or equal to ... we have to assume something about the reversibility of sh*t here.
\end{iquote}

\noindent This statement indicates a failure to connect a spontaneous heat flow with irreversible process in which, by definition, entropy increases. However, most students had a strong intuition that total entropy would increase, so considering the reversibility of the process does demonstrate some caution and awareness to check the isentropic edge case of the Second Law. 

Despite this question lacking any cues to think microscopically (\textit{i.e.} in terms of microstates or similar ideas) about this system, Beth, Chris, and Fred all brought up the idea that entropy was related to the number of states, a resource identified by Loverude which also showed up in the Partitioned Box question. For Chris, reasoning about the availability of microstates helped him rationalize his decision that the entropy of Block A decreased and the entropy of Block B decreases. For Beth and Fred, however, the resource did not seem to help since they struggled to think about what was happening to the blocks on the microscopic level. In Beth's case, it lead to the reasoning that the entropies of each block would not change since the number of states didn't change. She wrote:

\begin{iquote}
  \who{Beth [written]} For each block the number of states doesn't change, entropy doesn't change (?)
\end{iquote}

\noindent More on this resource will be discussed in Sec.~\ref{sec:EntropyMicrostates}.

\subsection{Question 3: Strings}

A detailed report on the student responses to the first section of this question (covering parts A, B, C, and D) was presented in a previous article~\cite{Crossette2020}, and an overview of student responses to the first three prompts is provided in Table~\ref{tab:stringspt1}. In summary, we observed that four out of the eight graduate students initially had a preference for an `intuitive' ranking of the probabilities of the three states saying that the more ``wiggly,'' ``wavy,'' or ``centered'' conformations of the string were more likely. These first two terms carry a similar ambiguity and connotation as `disorder,' though the idea of a ``centered'' string --- which came up in Alex and Fred's interviews --- carries more of a connotation of order, which could be similar to a finding that some students associate equilibrium with more order rather than disorder~\cite{Loverude2015}.

After considering the question on microstates and macrostates in part B, three students migrated to the correct answer that all three states are equally likely. In their consideration of macrostates of the string system, students generated a robust collection of appropriate macrostate classifications (see Table~\ref{tab:stringspt1}). We observed that the majority of students initially giving an unequal probability ranking in part A appeared to recycle their ranking metric as a way to classify macrostates. For example, Alex and Fred's metric of net deviation foreshadowed macrostates based on deviation, and Garth's idea of ranking probabilities by the distance between the ends of the string became his answer for part B. Additionally, three students indicated that the conformations themselves were macrostates. One student presented a formally correct argument that the particular arrangements of atoms and electrons in the string represented the microstates, but two others seemed to conflate macrostates with macroscopic objects. All together, these observations suggest some students may be projecting macrostate properties on to constituent microstates. 

On the whole, the interviewees had difficulty articulating the connection between the entropy of the string and the number of possible conformations of the string (part D), but four of the eight students were still able to reason about the connection productively. Furthermore, direct appeals to disorder were conspicuously missing from the students' discussion of this problem. Though the terms students did use were similarly qualitative and vague as `disorder', these observations suggest that graduate students have started to move beyond the ubiquitous mantra that entropy is disorder, at least when it comes to approaching practical situations.

In part E, we sought to see how well the students could connect the ideas about entropy and conformations to a more complicated system involving the strings, see Table~\ref{tab:stringspt2} for a summary of each student's responses. When considering what happens to the number of conformations of the strings, seven students (correctly) decided it would decrease when the circles and strings interact. The primary explanation for the decrease, given by six students, was that the circle particles would occupy space/exclude volume from the strings, which will limit their conformations. The seventh student, Erik, decided to recast the system into the more familiar context of an ideal gas expanding in to a vacuum, then claimed the decrease would occur transiently, as the circles created a sort of wind that would blow the strings into rightward pointing conformations.

\begin{iquote}
  \who{Erik} I would expect gale force winds in this direction [right] which would limit the number of conformations of the strings because all the strings will be flapping in that [right] direction...
  
  What do I expect to the possible conformations of the string when I remove this barrier and let the gas [circles] through: I expect it to decrease because all the strings are going to curve to the right.
\end{iquote}

Alex was the only student who did not explicitly state the number of conformations would decrease. Like Erik, he made a claim on the conformations themselves rather than the number of conformations. Alex invoked the idea of pressure, similar to Erik's idea of wind. When drawing concentration profiles, both Alex and Erik argued for flat, uniform profiles as would be the case without the strings, though Alex did briefly consider a profile with a dip in the middle region.

\begin{figure}
    \centering
    \includegraphics[width=.45\textwidth]{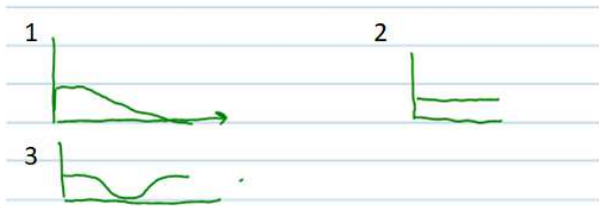}
    \caption{Three profiles drawn by Beth while answering the prompt about the final concentration profile. The numbers indicate the order in which the plots were drawn. When drawing the second plot, Beth was considering a case in which there were no strings. In the third sketch, Beth said she would not expect the concentration profile to go completely to zero in the middle region despite drawing it that way. \label{fig:BethProfiles}}
\end{figure}

On the whole, five students (not including Alex) decided the concentration profile would have a dip, which we considered the correct answer, similar to Beth's third sketch in Fig.~\ref{fig:BethProfiles}. Beth's sketches, coincidentally, make a good summary of the whole field of responses. Several students took a moment to consider a transient profile (similar to Beth's first sketch) early in the system's evolution despite no cue to do so. Two other students (Garth and Harry) explicitly considered the concentration profile for a system without the strings. Furthermore, as was mentioned previously, Alex and Erik argued for uniform profile matching Beth's second sketch.

A sketch drawn by Chris, however, was unique. Chris argued for a concentration profile of circles that was peaked in the middle region, and drew a profile which resembled a vertical reflection of the third sketch from Fig.~\ref{fig:BethProfiles}. He made an argument that the circles would get trapped in with the strings causing the flux of circles out of the middle region to be less than the flux into the region. This profile would result if the strings and circles could bind, releasing energy that would increase the overall entropy of the system.

\begin{table*}[htbp]
\renewcommand{\arraystretch}{1.1}
  \caption{Summary of graduate student responses to part E of the strings question. Lower case roman numerals in parentheses on title row refer to prompts in Fig.~\ref{fig:ChannelProfile}. For example sketches of the flat and dipped profiles drawn by students see Fig.~\ref{fig:BethProfiles}. \label{tab:stringspt2}}\vspace{-\baselineskip}
  \begin{center}
    \begin{tabularx}{\textwidth}{ccCcC}
    \hline
    \hline
      Student & Number of Conformations (i) & Reasoning (i) & Profile Drawing (ii) &  Entropy Connection\\ 
      \hline
      Alex      & deviate right  & pressure of circles `push' strings   & flat, briefly considers dip       & None (not prompted)     \\
      Daana     & decrease       & circles exclude volume               & dipped                            & spontaneous             \\
      Beth      & decrease       & circles exclude volume               & dipped                            & prompted                \\
      Erik      & decrease       & deviate right by pressure of circles & flat                              & prompted                \\
      Chris     & decrease       & circles exclude volume               & peaked                            & prompted                \\
      Fred      & decrease       & circles exclude volume               & dipped                            & spontaneous             \\
      Garth     & decrease       & circles exclude volume               & dipped                            & spontaneous             \\
      Harry     & decrease       & circles exclude volume               & dipped                            & spontaneous             \\
      \hline
      \hline
    \end{tabularx}
    \end{center}\vspace{-1.5\baselineskip}
    
\end{table*}

When asked to discuss what the decrease in configurations of the strings due to the presence of the circles (which Chris calls ``molecules'') meant for the entropy of the strings, Chris concisely articulated the relevant effect:

\begin{iquote}
  \who{Chris} The entropy of the strings has to go down because there's fewer states that they can access. But the entropy of the molecules goes way up because, you know, once I take that wall out, there's a ton more places the molecules could be.
\end{iquote}

\noindent This reasoning did not cause Chris to change his profile, however, despite one last question from the interviewer asking whether the peaked profile was his final answer.

Throughout the discussion of the question, seven students made a connection between the number of configurations and the entropy of the strings, though three of them did so only after a suggestive question from the interviewer. Alex did not make this connection, and Erik made the connection after some prompting, but still settled on a flat concentration profile.

Harry's discussion of the entropy trade-off between the circles and strings lead to a direct connection to a kind of ``force tending to push the balls out of the region'' with the strings. Indeed, there is an analogy between the change in entropy of the strings ($\Delta S$) associated per loss in volume ($\Delta V$) due to a circle which parallels the thermodynamic definition of pressure: $P = T \frac{\partial S}{\partial V}$.

This system with the strings, balls, and channel is also directly analogous to the first question on the interview with the partitioned boxes. The circle and square particles in the partitioned box question play similar roles as the circular particles and strings, respectively, in the channel system. In both systems the circular species are free to move throughout, while the square particles and strings are confined to one single region. The key difference between these two systems, however, is that the species in the channel question take up volume, unlike the ideal gases in the partitioned box system which are point-like. As discussed in Sec.~\ref{sec:InterviewQuestions}, the non-trivial volume of the circles and strings is what results in the entropic force causing the dip in the concentration of the circles in the middle region. If, instead, the circles were point-like and the strings were line-like, the circles would have a uniform concentration profile, similar to how the circle ideal gas particles in the partitioned box question equally distribute themselves between the two halves of the box despite the presence of the square particles.

The vast majority of graduate students picked up on the fact that the circles in the strings question took up volume, which lead to a non-trivial concentration profile. This is noteworthy, especially since students also were able, for the most part, to reason correctly about the ideal gases in the first question, which were also drawn (misleadingly) with non-zero volume. None of the students recognized the connection between these two questions, though Chris and Daana explicitly asked the interviewer whether they should consider the 2D area of the circles as they were drawn. Nonetheless, the ability of the graduate students, as a whole, to recognize the most salient features of each question and reason accordingly given the different situations demonstrates well-developed physical reasoning skills.

\subsection{Question 4: Expansions}

\newcommand{\ra}{$\rightarrow$}
\newcommand{\dSiso}{\Delta S_{\text{iso}}}
\newcommand{\dSFE}{\Delta S_{\text{FE}}}

The dependencies of entropy on volume and temperature have emerged as two themes woven through this set of interview questions. While the preceding tasks only required students to think about one dependence at a time, this task required students to contend with both the volumetric and thermal aspects simultaneously. Several students commented on recognizing either the isothermal or free expansion in a prior context:

\begin{iquote}
  \who{Beth} I'm really, really trying to remember... I feel like I've talked about this exact problem in a class before, and I remember it being a very obvious thing that I just don't remember
\end{iquote}

\noindent Despite the students' familiarity with the expansions, this task proved to be the most challenging of the interview. Students struggled to form sound arguments, they expressed less confidence in their answers, and their responses were largely idiosyncratic with little consistency from student to student.

In their study exploring this task, Bucy \etal\ gave this prompt to students both at the beginning and end of an upper-division thermal physics course. Generally, they found students relied heavily on mathematical equations to approach this task. On the post-test, they found that undergraduates applied the thermodynamic definition of entropy to correctly identify the sign of $\Delta S$ in the isothermal expansion, but struggled with determining the sign in the free expansion. Many students on the post test made a claim that the temperature of the gas would decrease, despite the prompt explicitly stating the temperature was constant. On the comparison task, they saw only two (out of seven) students correctly identify that the changes in entropy were identical between the two processes on the post-test, with only these two students explicitly using entropy's state function property.

\begin{table*}[htbp]
\renewcommand{\arraystretch}{1.1}
  \caption{Summary of graduate student responses to the expansions question. In this table, the quantities $\dSiso$ and $\dSFE$ represent the magnitude of the change in entropy. $\Delta T_{\text{FE}}$ is the change in temperature of the gas undergoing the free expansion. A ``yes'' in this column indicates the student considered this at any point while reasoning about the task, regardless of their final conclusion. \label{tab:expansions}}\vspace{-\baselineskip}
  \begin{center}
    \begin{tabularx}{\textwidth}{cCCCc}
    \hline
    \hline
      Student & Sign $\dSiso$ & Sign $\dSFE$ & Entropy Change Comparison & Claimed $\Delta T_{\text{FE}} < 0$ \\ 
      \hline
      Alex      & +                   &  +~\ra~-~\ra~0~\ra~+ & $\dSiso < \dSFE$  & yes   \\
      Daana     & +                   & 0~\ra~+              & $\dSiso = \dSFE$  & no  \\
      Beth      & 0                   & -                    & $\dSiso < \dSFE$  & yes    \\
      Erik      & 0~\ra~+             & +~\ra~0~\ra~+~\ra~0   & $\dSiso > \dSFE$\footnote{inferred from final answers to prior questions} & yes    \\
      Chris     & +                   & 0~\ra~+              & $\dSiso > \dSFE$  & yes    \\
      Fred      & +                   & +                    & $\dSiso < \dSFE$  & no    \\
      Garth     & +                   & 0~\ra~+              & $\dSiso = \dSFE$  & yes    \\
      Harry     & +                   & 0                    & $\dSiso > \dSFE$  & no    \\
      \hline
      \hline
    \end{tabularx}
    \end{center}\vspace{-1.25\baselineskip}
    
\end{table*}

Our study found very similar trends with how graduate students responded to this question. On the first task of determining the sign of the change in entropy, $\dSiso$, of the gas undergoing the isothermal expansion, most students reasoned it would be positive. However, only two students were able to both activate and successfully reason by using the thermodynamic definition of entropy: $\Delta S = Q/T$. Most other students relied on qualitative intuitions on how entropy is related to volume and temperature. The intuition that entropy depended on temperature lead Beth to her answer of $\dSiso = 0$ (since the temperature was constant). Erik's initial response of $\dSiso = 0$ came from recognizing this expansion as a reversible process, where $\Delta S_{\text{tot}} = 0$, and thinking this meant that $\dSiso$ had to be zero (he later corrected this response).

Again mirroring Bucy's findings, the graduate students we interviewed struggled more with determining the sign of $\dSFE$, the change in entropy of gas undergoing the free expansion. This is somewhat surprising, since the free expansion is a canonical example of an irreversible process (in which entropy must increase), but six out of the eight graduate students interviewed at one point in the interview stated that $\dSFE$ was zero. Two students even considered that it was negative due to an intuition that the gas would cool and that entropy was related to temperature.

Three students (Garth, Harry, and Daana) considered that $\dSFE$ was zero after realizing that the change in energy of the gas was zero. Daana and Harry, the only two students to explicitly invoke the equation $\Delta S = Q/T$ in the prior part, incorrectly applied it to the free expansion (though Daana later corrected). This could be due to an implicit assumption that the process could be treated as quasistatic where the relationship equating the change in entropy to heat flow is true, a possible artefact of a heavy emphasis on quasistatic processes in many thermal physics courses.

Garth reasoned about the free expansion more conceptually by thinking about how changes in the volume and temperature would affect the entropy

\begin{iquote}
  \who{Garth} Although the gas has more volume, it's temperature is decreasing.
\end{iquote}

\noindent This intuition that an expanding gas must cool was brought up by most of the students at some point while engaging with this question, and five brought it up while thinking specifically about the free expansion. Chris justified this intuition with a real-world experience

\begin{iquote}
\who{Chris} It ends up colder, I guess, by the gas law. In the same way that your propane gets cold when you turn on your stove.
\end{iquote}

\noindent The authors admit to having a similar, visceral sense for an expanding gas to cool. While a free expansion is an idealization and special case that can not truly happen in nature, it helps to illustrate the cause for temperature changes of expanding and contracting gases: work on the environment. However, it should be noted that in the case of a propane tank releasing a gas, the primary effect causing the tank to cool is the latent heat of the endothermic phase transition of liquid propane vaporizing.

Only Daana, Erik, and Garth explicitly identified the free expansion as an irreversible process. For Daana and Garth, this realization seemed to unlock the whole problem since it preceded them correcting their answers for the sign of $\dSFE$ and correctly comparing the changes in entropy of the two processes. In Erik's case, it did not provide sufficient confidence to finalize his choice of sign, and he went on to change his choice multiple times before finally settling on $\dSFE = 0$.

As can be seen in Table~\ref{tab:expansions}, there was very little consensus when it came to comparing the magnitudes of $\dSiso$ and $\dSFE$. In addition to recognizing the free expansion as an irreversible process, Daana and Garth were the only two students who recognized that the two gases started and ended in the same initial and final states, respectively. No students explicitly mentioned that entropy was a state function, though Daana and Garth demonstrated an understanding of the salient reasoning. For the students saying that $\dSiso$ was greater, most arrived at the conclusion as a result of deciding that $\dSFE$ was zero. Alex and Fred settled on $\dSFE$ as being greater with both mentioning the ability of the gas to exchange energy with, and lose energy to, the environment. For Beth, the magnitude of $\dSFE$ was greater simply as the logical conclusion of deciding that $\dSiso$ was zero.

Students fared better on the question about the changes in entropy of the surrounding environment. All students recognized that the environment outside of the thermally isolated free expansion would be zero since it was decoupled from the gas and no heat could flow. Only six students, though, concluded that the entropy of the environment outside the gas isothermally expanding was negative. Fred claimed that the entropy of the environment would increase, and Erik struggled separating the changes in entropy of the gas, outer environment, and universe, a difficulty also found in prior studies~\cite{Christensen2009, BucyPERC2006}. 

\section{Discussion of Inter-Question Themes and Resources}
\label{sec:Resources}

In the preceding sections, we sought to narrowly and deeply examine student responses to each question in our interview independently of the other tasks on the interview questionnaire. In this section, we will take a step back to look more broadly at themes that emerge when considering the interview as a whole. In examining common elements of reasoning employed by students across questions, we will identify student ideas that can be categorized as conceptual resources, as introduced by Hammer~\cite{Hammer}.

According to Hammer, a conceptual resource is analogous to a block of computer code that performs a specific task, and is often pasted into new code without any thought given to the inner workings of the block. In parallel to this computer science metaphor, empirically, students use conceptual resources as tools, often times without justification for why the tool is appropriate or correct. Consequently, a particular resource may be correct or appropriate in one context, but incorrect or inappropriate in a different context. From an instructional perspective, an understanding of the resources students use to think about the concepts used and taught in classrooms helps an instructor better relate and communicate with students. To conclude this section, we will compare the resources students used to answer the interview prompts with the ways they reported conceptualizing of entropy.

\subsection{Entropy is related to the number of (micro)states}
\label{sec:EntropyMicrostates}

At least one student employed the reasoning that entropy is related to a number of states in every single question on the interview, even in the Blocks and Expansions questions where there was neither a cue nor a need to think about microscopic quantities. In his study of undergraduate student reasoning about entropy and equilibrium, Loverude identified a resource he labeled as ``entropy is related to multiplicity''~\cite{Loverude2015}. Though we believe these two resources to be nearly identical, we chose to refer to this reasoning with language that reflected the phrasing used by students. Students talked about a number of states or a number of microstates much more frequently than they mentioned the word `multiplicity.'

Additionally, Bucy noticed that students have a preference for the microscopic/statistical definition of entropy over the macroscopic/thermodynamic definition~\cite{BucyPERC2006} which is consistent with students in our interview thinking about microstates in the Blocks and Expansions questions. One possible reason for this affinity stems from the statistical definition of entropy
\begin{equation}
S = -k_B \displaystyle\sum_i p_i \ln p_i
\label{eq:generalEntropy}
\end{equation}

\noindent which provides a more \textit{a priori} conceptualization of entropy without relationship to other quantities. Incidentally, students in this study seemed to invoke the general form above in Eq.~\ref{eq:generalEntropy} about as often as the standard form, Eq.~\ref{eq:BoltzmannEntropy}. Furthermore, this definition holds true more generally than the thermodynamic definition of entropy ($\Delta S = Q/T$) which does not apply to processes in which entropy changes, but the heat flow is zero. 

Many students used this resource appropriately and productively. In the Partitioned Box question, Daana invoked Eq.~\ref{eq:generalEntropy} and Garth thought about the number of ways to split up the particles with combinations and binomial coefficients to reason about the entropy of each of the four states. In the expansions question these same two students utilized reasoning about microstates and configurations, respectively, to decide that the entropy of the gas expanding freely increased despite the heat flow being zero.

In the Strings question, half the students made a substantive connection between the configurations of a string and the entropy of the string. Harry even made a deeper connection between the entropy of the string and the density of states:

\begin{iquote}
  \who{Harry} So there's only one microstate that corresponds to this string reaching as far as it can. And then this one, there's a lot more, um, a much higher density of microstates that lead to the string reaching that far because it can be like this, or this....
\end{iquote}

As mentioned in Sec.~\ref{sec:Q2_Blocks}, Chris effectively used microstates to reason about the changes in entropy of the two blocks.

\begin{iquote}
  \who{Chris} I suppose the entropy of the higher temperature one will decrease because it has to give up some of its energy to solid B and the entropy of solid B will increase because it's sort of received some energy has more microstates allowed because there's more momentum possible in the little particles. But it will increase more than the decrease in entropy of solid A, so that the total entropy is... at the end is bigger than at the beginning.
\end{iquote}

\noindent However, Fred and Beth also brought up this resource in the Blocks question, but it did not help them reason productively. Fred said the entropy of both blocks would increase. Beth thought the entropy of the two block system would remain constant because she could not see how the total number of states would change. A tendency for students to treat entropy as a conserved quantity as been noted in other studies~\cite{Leinonen2015, Christensen2009, Bennett2007}. In this case with Beth, it seems like the idea of constant entropy was due to a sense that the total space of states possible to the system (independent of energy constraints) remains static, with only the number of \textit{accessible} states changing depending on the macrostate of the system. Her reasoning also dovetailed with the relationship between entropy and temperature will be discussed next.

\begin{iquote}
  \who{Beth} I remember $S$ being the sum of states. I think there was a log in there somewhere, but whatever... Yeah, I think that was log of the microstates. Um, yeah, so like naively, the way we think about it in condensed matter is the number of states is related to the temperature because the more things can move the more states they have, so higher temperature would be more possible states and lower temperature means less... states.
\end{iquote}

\subsection{Entropy/Temperature association}

The second conceptual resource we observed students use frequently was an association between entropy and temperature, as if a monotonic relationship existed between the two quantities. This association did not appear in the Partitioned Box question (which may not be surprising since temperature was not a salient feature of the task), but did appear in the three other physics-content tasks. Since such an association can be problematic in some circumstances, we have included an Appendix with a deeper discussion and counterexample showing a case were this association can fail, which might serve as a tool to help instructors who may want to address this association with students.

We noticed two primary use cases for this resource. In the first, students used temperature to directly compare the entropy of two different objects, as Fred does in this quote from the Expansions task:

\begin{iquote}
  \who{Fred} I would think that the one [gas] with the higher kinetic energy has more entropy, um, just cuz of my, like elementary understanding entropy as being like chaos. I think of the rapidly moving molecules as having more entropy.
\end{iquote}

\noindent On the surface, such reasoning is valid if temperature is the \textit{only} difference between two objects. However, by making it seem like changing an object's temperature was the only way to change it's entropy, this reasoning did lead some students astray as it did for Beth in the Expansions task:

\begin{iquote}
  \who{Beth} Also, vaguely, entropy has something to do with temperature. I don't remember what, there was a formula I learned at some point, but if temperature is not changing then, naively, the entropy shouldn't change either. Maybe.
\end{iquote}

\noindent This quote from Beth also demonstrates the second way in which students used this resource: to think about how changes in temperature affect the entropy of a single object.

As already discussed in Sec.~\ref{sec:Q2_Blocks}, students also used this resource heavily in the Blocks question. But somewhat surprisingly, this resource also briefly appeared in the Strings question. Erik brought up this resource when thinking about the changes in entropy of the circles and strings as the circles diffused through the channel. He seemed to wonder whether the circles being at some temperature might impart some energy to the strings and increase the entropy of the strings:

\begin{iquote}
  \who{Interviewer} So, does that [having a circle in the strings region] do anything to the entropy of the strings?
  
  \who{Erik} The entropy of the strings itself? Okay, let me think about it this way... If, if I had a very hot gas here and nothing in here, and the strings were here, I let this loose, the temperature of this, the strings, must come up. So, I'm guessing the entropy goes up?? But the entropy is not related to the temperature directly...
\end{iquote}

After a brief discussion, Erik did concluded that the entropy of the strings would decrease due to the decrease in number of conformations. Two other students, Harry and Chris, also worried about how the temperature of the water bath or circles might effect the strings, though more tangentially.

\subsection{Other qualitative entropy proxies}

Students used a variety of qualitative terms to conceptually describe entropy, some were distinct enough to rise to the status of resource. One such intuition was that entropy was related mixing. This occurred most commonly in the Partitioned Box question, but also appeared in the Strings question when Alex talked about how the number of conformations of the string would relate to the entropy of the string. In Tab.~\ref{tab:resources}, we outline the elements of reasoning we believe reach the status of conceptual resource.

Other terms which, on their own, did not appear as consistently may not be particularly interesting, but when considered together the collection of terms tells a bigger story. In the Partitioned Box question, many students described entropy as something related to the ``distribution'' of particles. In the Strings question, students seemed to favor terms like ``chaotic,'' ``squiggly,'' and ``wavy.'' In the Blocks and Expansions questions, students leaned on an association between entropy and temperature. The variety of ideas used to describe entropy suggest that graduate students have flexible conceptions of entropy that they adjust when approaching different situations.

\newcolumntype{R}{>{\raggedright\arraybackslash}X}
\begin{table*}[htbp]
\renewcommand{\arraystretch}{1.1}
  \caption{Prevalent resources used by graduate students in this interview and student quotes exemplifying each resource. The first three resources were identified by Loverude~\cite{Loverude2015} and also appeared in our study. The second three are new resources we identified. The second column gives the question from which the student quote came. \label{tab:resources}}\vspace{-\baselineskip}
  \begin{center}
    \begin{tabularx}{\textwidth}{lcR}
    \hline
    \hline
      Resource & Question & Example Student Quote \\ 
      \hline
      Entropy is related to the number of microstates & Strings         & \textit{I related the entropy to the number of microstates}          \\
      Entropy of a system must always increase        & Blocks          & \textit{The entropy of the entire system... has got to increase}               \\
      Entropy is related to disorder                  & Strings         & \textit{You expect to find it in a higher entropy state where things are disordered}          \\
      \hline
      Entropy is related to temperature               & Blocks          & \textit{I'm definitely associating entropy with temperature}           \\
      Entropy is related to mixing                    & Partitioned Box & \textit{The system with the highest entropy is... the system that is most well mixed.}          \\
      Entropy is related to information               & Partitioned Box & \textit{The entropy should be the lowest... because we have more information}         \\
      \hline
      \hline
    \end{tabularx}
    \end{center}\vspace{-1.5\baselineskip}
    
\end{table*}

\subsection{Students' self-reported conceptualizations of entropy}

Two students had more unique resources that they brought up multiple times in their interviews. First, Alex had a strong affinity for thinking of entropy as information, and spontaneously brought up the term in the Partitioned Box and Expansions questions.

\begin{iquote}
  \who{Alex} Entropy itself is directly related to information by the Shannon Entropy.
\end{iquote}

\noindent This proxy did not appear to help in the Partitioned Box task since Alex settled on an incorrect entropy ranking, but this intuition did appear to help him realize that the entropy of the freely expanding gas increased in the Expansions question. 

The second student, Fred, had an affinity for the term chaos, bringing it up spontaneously in the Blocks, Strings, and Expansions questions. However, despite raising this association in the Strings question, Fred defied this intuition. He ranked the probabilities of the arrangements as $P(a)>P(c)>P(b)$ based on an expectation that strings with less deviation from the center were more likely. Furthermore, he admitted to thinking that the \textit{entropy} of the three configurations were the same despite his intuition that the more ``squiggly'' strings would have more entropy. In terms of the resources in Tab.~\ref{tab:resources}, we classified this chaos resource under the `Entropy is related to disorder' resource.


A consensus from our interviews with thermodynamics instructors lead us to expect that when asked about entropy students would default to the mantra that entropy is disorder. Indeed, students brought up the idea of disorder, but much less than expected. On the four physics content questions, Beth used the term in the Partitioned Box and Strings questions, Garth used the term in the Blocks question, and Erik used the term ``ordered'' in the Partitioned Box question. Two additional students, Daana and Fred, brought up disorder when directly asked how they conceptualized entropy (see next section). Exhaustively, that is a total of five students mentioning either disorder or order in six different situations. This observation reflects the findings of Leinonen's study which found that few undergraduates used the association of entropy with disorder~\cite{Leinonen2015}.

To close the interviews, we asked students a fifth question: how they personally conceptualized entropy. As a proportion of the total time, the amount of discussion on students conceptualization of entropy was much less than any single content question, but was the most common place for students to bring up the association between entropy and disorder. When taken with the rest of the results of the interviews, this suggests that graduate students have a robust and somewhat varied set of tools with which to practically reason about entropy, but still have trouble conceptualizing the quantity abstractly.

The most common response, given by six students, in some way mentioned that entropy was related to the number of configurations or (micro)states. The two remaining students, Alex and Fred, mentioned information and disorder, respectively. In addition to Fred, only Daana brought up that entropy was related to disorder on this question, but went on to say that disorder was an incomplete way of thinking about entropy. Three students, Alex, Beth, and Erik, admitted to struggling with conceptualizing entropy.


Alex's unique association between entropy and information also seemed to have a connection to prior research experience. When explaining his conceptualization, Alex discussed how it applied in the Ising magnet model, a system with which he had prior research experience. Garth also had an somewhat unique association of entropy as ``the arrow of time'' since systems always evolve towards higher entropy, though Garth's association could be a manifestation of the ``entropy of a system must always increase'' resource. Fred had a similar association of entropy as something that always increases or stays the same.

Overall, the strong preference for conceptually relating entropy to the multiplicity of microstates mirrors a finding from Bucy's study that finds undergraduates favored the statistical definition of entropy~\cite{BucyPERC2006}. This stated preference for microstates is consistent with our pool of students frequently thinking about microstates while solving problems.

\vspace{-.5\baselineskip}
\section{Conclusion}
\label{Sec:Conclusion}

We interviewed eight students in the physics graduate program at the University of Colorado Boulder. The interview consisted of thermal physics tasks designed to probe students' understanding of and practical reasoning skills with entropy. The questions dealt with entropy from the microscopic and macroscopic perspectives, ideal gases, and a novel biophysical context involving a system with a dynamic string. We presented student responses on a question-by-question basis, then addressed themes and conceptual resources which emerged throughout the interview.

In the Partitioned Box question about two ideal gases mixing across a membrane, we found that some students had a preference for thinking about entropy by relating it to mixing. Students were largely able to correctly identify the state with the most entropy, despite some dissonance regarding the unequal pressures in the two halves of the box in the maximum entropy state.

Students also demonstrated a strong association between entropy and temperature despite no such explicit, mathematical relationship existing. This association presented itself throughout the interview, but featured most prominently in the Blocks question. In the Appendix, we present a counterexample that may aid instructors in addressing problematic aspects of this association.

On the Strings question, students: seemed to project macrostate properties onto constituent microstates; were largely able to rank equally the probability of microstates; and struggled more with connecting multiplicity with entropy in this novel context. A more detailed report of the first part of this question was presented in a prior publication~\cite{Crossette2020}. When considering the entropic interaction between strings and another species, students had more success in connecting multiplicity to entropy. 

Graduate students generally did not consider the state function property of entropy in the Expansions question, a task taken from Bucy~\cite{BucyPERC2006}. Students had an easier time determining the sign of the change in entropy of the quasistatic, isothermal expansion than the free expansion. Overall, there were many similarities between graduate student responses in our interviews and undergraduate responses in Bucy's study.

Throughout the interview, students frequently used a conceptual resource, `entropy is related to the number of microstates,' a resource also identified by Loverude~\cite{Loverude2015}. Additionally, when asked directly, many graduate students explicitly stated that they conceptualized entropy as being related to the number of states. This consistency between the practical use of the resource in reasoning and the students' metacognitive conceptualization of entropy demonstrates a high level of awareness indicative of more advanced physicists~\cite{Irving2013}. This is in contrast with the resource relating entropy with temperature. Students frequently used this resource practically on the interview tasks, but did not mention it when explicitly asked how they conceptualize entropy.

Explicit connections between entropy and disorder, another common connection for undergraduates identified by Loverude~\cite{Loverude2015}, were less common in our interviews, which echoes a study by Leinonen which noted an absence of the term ``disorder'' appearing among undergraduates~\cite{Leinonen2015}. Additionally, a third entropy-related resource, `entropy of a system must increase,' identified by Loverude appear in our interviews. We also identified new entropy related resources relating entropy to temperature, mixing, and information. Awareness of the resources commonly used by students can help instructors communicate students more effectively. In particular, instructors may wish to directly address the subtleties of the relationship between entropy and temperature with their students, or make use of students' preference for thinking about entropy in terms of microstates. Future work with undergraduate students will better inform and expand our instructional recommendations.

This work is limited by the small sample size and the selection effect created by the solicitation of participants. Additionally, this study occurred at a large, R1 university with a selective graduate physics program which introduces another layer of selection effects. Thus, the resources we identify are not likely to be exhaustive or representative of all the resources commonly used by graduate students. Future work will compare the responses of undergraduate and graduate students to the same interview using the same coding scheme. This direct comparison will allow the identification of persistent conceptual difficulties with entropy faced by physics students, and also expand the scope of this work to include undergraduate students who may pursue careers outside of academia.

\vspace{-.5\baselineskip} 
\section*{Acknowledgements}
This work was funded by the CU Physics Department. A special thanks to the student participants, the faculty who reviewed the interview protocol, and the members of PER@C group for their help refining the interview questions.

\section*{Appendix: Counterexample for monotonic relationship between Entropy and Temperature}

In Sec.~\ref{sec:Q2_Blocks} we observed many students using a resource that increasing temperature increases entropy, as if
entropy was a strictly monotonically increasing function of temperature. In many cases, such an intuition is correct and valid reasoning, yet no such general mathematical relationship exists between temperature and entropy in the context of thermal physics. The closest equations are the Sackur-Tetrode equation for ideal gasses and the high/low temperature limiting cases for the entropy of an Einstein solid, all of which relate entropy to \textit{energy}, not temperature. However, energy and temperature are directly related for ideal gasses, and the energy as a function of temperature for Einstein solids takes the form:
$$ U(T) = \varepsilon N \left(\frac{1}{2} + \frac{1}{e^{\varepsilon/kT} - 1} \right) $$

\noindent where $N$ is the number of oscillators and $\varepsilon$ is the harmonic oscillator energy level spacing. Since this is a monotonically increasing function of temperature and the entropy equations for Einstein solids are monotonically increasing functions of energy (within their realm of applicability), entropy of an Einstein solid (and an ideal gas) \textit{must be} a monotonically increasing function of temperature \textit{when all other thermodynamic variables are held constant}.

This is where the resource can break down. Thermodynamic systems, like ideal gases, typically have multiple thermodynamic variables that can change independently of each other, so an increase in entropy due to an increase in temperature can be offset by a change in some other thermodynamic variable. In the following proof, we will layout a process in which temperature decreases and entropy increases.

\begin{figure*}
    \centering
    \includegraphics[width=\textwidth]{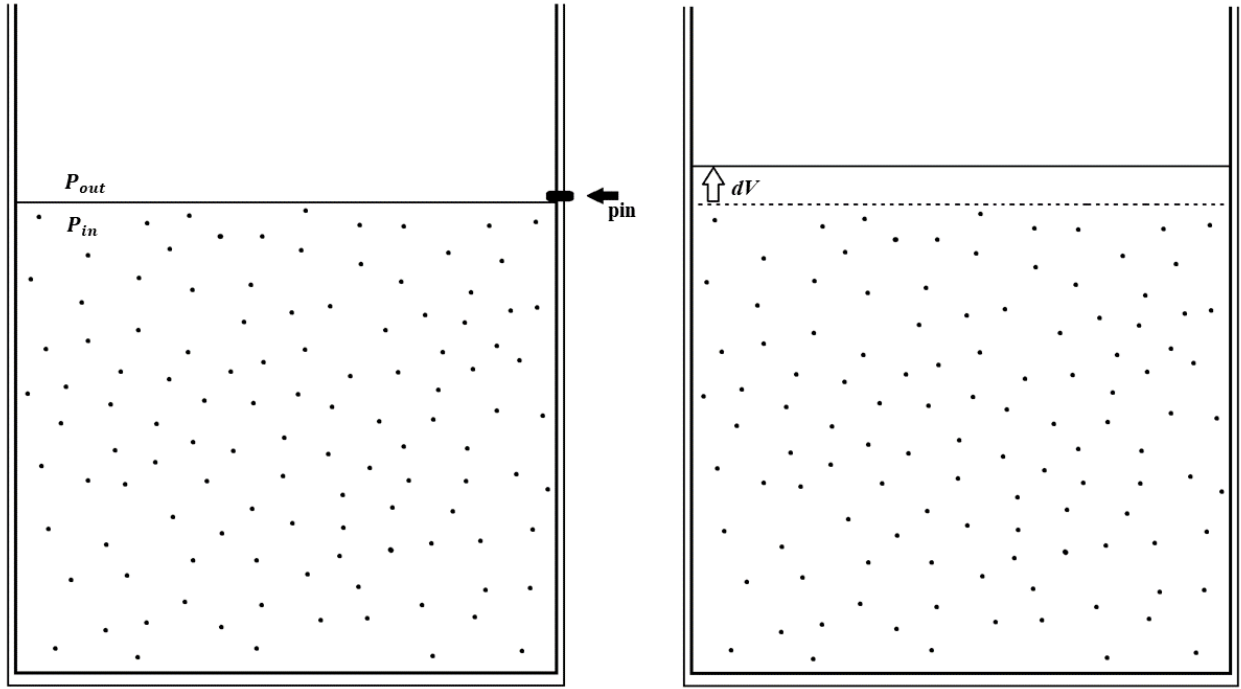}
    \caption{A hot gas at a higher pressure than the outer surroundings inside a piston. A pin holds the piston in place. When the pin is removed, the gas begins a spontaneous expansion. An infinitesimal change in volume is shown on the right half of the figure. \label{fig:CE_expansion}}
\end{figure*}

\renewcommand\qedsymbol{$\blacksquare$}

\begin{proof}
Consider a gas inside a sealed, thermally insulated piston like the system depicted in Fig.~\ref{fig:CE_expansion}. The pressure inside the piston is much greater than the pressure outside the piston ($P_{in} >>> P_{out}$). Initially, the top of the piston is held in place by a pin. We will consider two processes. First, the pin will be removed and the gas will be allowed to expand quasi-statically from the initial volume, $V_i$, to a final volume of $V_i + dV$. In the second, the pin will be removed and the gas will expand non-quasistatically from $V_i$ to the same final volume of $V_i + dV$.

In the quasi-static expansion, we can invoke the thermodynamic identity to consider the change in entropy:

\begin{equation}
\begin{aligned}
dU	=& ~T dS - P dV \\
dS  =& ~\frac{dU}{T} + \frac{P}{T} dV
\end{aligned}
\label{eq:thermoID}
\end{equation}

In this form, we have the change in entropy in terms of the change in energy $dU$ and the change in volume, $dV$. We want to consider the relationship between changes in entropy and changes in temperature, so we need to replace the $dU$ term with a $dT$ term. From the equipartition theorem ($U =\frac{3}{2} N k_b T$), we have a direct relationship between the energy and temperature of the gas. This will let us to change variables by replacing the $dU$'s with $dT$'s:

\begin{equation}
\begin{aligned}
dU 	=& ~\frac{3}{2} Nk_b dT
\end{aligned}
\end{equation}

Substituting this in to Eq.~\ref{eq:thermoID} and cleaning up:

\begin{equation}
\begin{aligned}
dS  =& ~\frac{3}{2} \frac{N k_b}{T} dT + \frac{P}{T} dV
\end{aligned}
\label{eq:diffEntropy}
\end{equation}

Since the gas is insulated, we know that the heat flow, $Q$, is zero, so only work being done on the piston will change the energy of the gas. We can use this to relate $dU$, and by extension $dT$, to the change in volume $dV$:

\begin{equation}
\begin{aligned}
dU  =& ~\cancelto{0}{Q} + W   \\
dU  =& ~-P dV \\
\frac{3}{2} Nk_b dT =& ~-P dV
\end{aligned}
\label{eq:EnergyChange}
\end{equation}

Use this relationship to substitute into the first term of Eq.~\ref{eq:diffEntropy} and we will see that the change in entropy of the gas is zero:

\begin{equation}
\begin{aligned}
dS  =& ~ -\frac{P}{T}dV  + \frac{P}{T} dV\\
dS  =& ~ 0
\end{aligned}
\label{eq:dS_quasistatic}
\end{equation}

This result should not be a surprise, since all quasi-static, adiabatic processes are isentropic. In this case, temperature decreased, but entropy remained constant which \textit{is} consistent with a monotonic relationship between temperature and entropy. Now let's consider the non-quasistatic expansion.

We have chosen a situation in which $P_{in} >>> P_{out}$, so that when the pin is removed the top of the piston will accelerate quickly (to a speed faster than the speed of sound inside the piston) and create a layer of pressure lower than the average pressure of the gas just beneath the top of the piston. This region is depicted on the right side, between the dashed and solid lines, in Fig.~\ref{fig:CE_expansion}.

This region of lower pressure will cause the gas to do \textit{less} work than the $PdV$ from the quasi-static case. This can be understood microscopically as fewer gas particles colliding with the piston and, overall, losing less energy than in the quasi-static case.

In the two processes, the gas expands to the same final volume, but ends up with a higher internal energy (and therefore, temperature) in the non-quasistatic case. We can use the Sackur-Tetrode equation to compare the final entropies of the gas in the two processes.
\begin{equation}
\frac{S}{Nk_b} = \ln \left[ \frac{V}{N} \left( \frac{4\pi m}{3 h^3} \frac{U}{N}\right)^{3/2}\right] +5/2
\end{equation}

Since the gas undergoing the non-quasistatic process ends at the same volume as the gas in the quasistatic process, but at a higher energy, we can see that $S_{f,nqs}$ is greater than $S_{f,qs}$. Also, the initial entropies in both cases are identical, because the two processes start in the same initial state. Therefore, $\Delta S_{nqs}$ must be greater than $\Delta S_{qs}$, and since $\Delta S_{qs}$ was zero, $\Delta S_{nqs}$ must be positive.

Therefore, we have that in the non-quasistatic case temperature \textit{decreases} and entropy \textit{increases}.

\end{proof}

This result can be understood qualitatively as well. When $P_{in} >>> P_{out}$, the expansion of the gas is a spontaneous, non-reversible process meaning entropy must increase. The gas loses energy, therefore it's temperature decreases. The increase in entropy due to the volume expanding, then, must be larger than the decrease in entropy due to the temperature decrease.

Somewhat ironically, while disproving a strictly monotonic relationship between temperature and entropy, the intuition that higher temperature means more entropy is vindicated to some extent. In the comparison of the final entropies of the gasses in the two processes, we found that the higher temperature gas (in the non-quasistatic expansion) had the larger entropy. This demonstrates that the association can be useful and productive, when it is applied correctly. The resource is valid when temperature is the \textit{only} difference between two systems.

We hope that this discussion may prove to be useful to instructors and students looking to better teach and understand the subtleties of the connection between temperature, energy, and entropy. This example could serve as the basis for homework questions or lecture material. For an additional exercise, one can also show that in a non-quasistatic compression of a gas (the inverse of non-quasistatic expansion), entropy can \textit{decrease} while temperature \textit{increases}.

\begin{thebibliography}{99}


\bibitem{Dreyfus2015}
B.W. Dreyfus, B.D. Geller, D.E. Meltzer, and V. Sawtelle, Resource Letter TTSM-1: Teaching Thermodynamics and Statistical Mechanics in Introductory Physics, Chemistry, and Biology, Am. J. Phys., \textbf{83}, 1 (2015)

\bibitem{BucyPERC2006}
B.R. Bucy, J.R. Thompson, and D.B. Mountcastle, What Is Entropy? Advanced Undergraduate Performance Comparing Ideal Gas Processes, AIP Conference Proceedings, \textbf{818}, 1 (2006).

\bibitem{Smith2015}
T.I. Smith, W.M. Christensen, D.B. Mountcastle, and J.R. Thompson, Identifying student difficulties with entropy, heat engines, and the Carnot Cycle, Phys. Rev. ST Phys. Educ. Res., \textbf{11}, 2 (2015)

\bibitem{Loverude2015}
M. Loverude, Identifying student resources in reasoning about entropy and the approach to thermal equilibrium, Phys. Rev. ST Phys. Educ. Res., \textbf{11}, 2 (2015)


\bibitem{Leinonen2015}
R. Leinonen, M.A. Asikainen, and P.E. Hirvonen, Grasping the second law of thermodynamics at university: The consistency of macroscopic and microscopic explanations, Phys. Rev. ST Phys. Educ. Res., \textbf{11}, 2 (2015)

\bibitem{Christensen2009}
W.M. Christensen, D.E. Meltzer, and C.A. Ogilvie, Student ideas regarding entropy and the second law of thermodynamics in an introductory physics course. Am. J. Phys., \textbf{77}, 907 (2009)

\bibitem{Carson2002}
E.M. Carson and J.R. Watson, Undergraduate students' understandings of entropy and Gibbs free energy, U. Chem. Educ., \textbf{6}, (2002)

\bibitem{Loverude2010}
M. E. Loverude, Investigating student understanding for a statistical analysis of two thermally interacting solids, Proceedings of the 2010 Physics Education Research Conference, Portland, 2010 (American Association of Physics Teachers, American Institute of Physics, Melville, NY, 2010).

\bibitem{Bennett2007}
M. S{\"o}zbilir and J. M. Bennett, A study of Turkish chemistry undergraduates' understanding of entropy, J. Chem. Educ. \textbf{84}, 1204 (2007).

\bibitem{Haglund2015}
J. Haglund, S. Andersson, and M. Elmgren, Chemical engineering students' ideas of entropy, Chem. Educ. Res. Pract., \textbf{16}, 537 (2015)

\bibitem{Geller2014}
B. D. Geller, B.W. Dreyfus, J. Gouvea, V. Sawtelle, C. Turpen, and E. F. Redish, Entropy and spontaneity in an introductory physics course for life science students, Am. J. Phys. \textbf{82}, 394 (2014).

\bibitem{Crossette2020}
N. Crossette, M. Vignal, and B. R. Wilcox, Investigating how graduate students connect microstates and macrostates with entropy, 2020 PERC Proceedings (2020)

\bibitem{thinkaloud}
M. E. Fonteyn, B. Kuipers, and S. J. Grobe, A description of think aloud method and protocol analysis. Qualitative health research \textbf{3}, 4, 430-441 (1993)

\bibitem{Hammer}
D. Hammer, Student resources for learning introductory physics, Am. J. Phys. \textbf{68}, (2000)

\bibitem{BucyThesis}
B. R. Bucy, Ph.D. thesis, University of Maine, 2007.

\bibitem{Kautz-Macro2005}
C.H. Kautz, P. R. L. Heron, M. E. Loverude, and L.C. McDermott, Student understanding of the ideal gas law, Part I: A macroscopic perspective, Am. J. Phys. \textbf{73}, 11, (2005)

\bibitem{Kautz-Micro2005}
C.H. Kautz, P. R. L. Heron, P. S. Shaffer, and L.C. McDermott, Student understanding of the ideal gas law, Part II: A microscopic perspective, Am. J. Phys. \textbf{73}, 11, (2005)

\bibitem{Irving2013}
P. W. Irving and E. C. Sayre, Physics identity development: A snapshot of the stages of development of upper-level physics students, Journal of the Scholarship of Teaching and Learning, \textbf{13}, 4, 68-84, (2013)

\end{thebibliography}
\end{document}